\documentclass[aps,pra,amssymb,twocolumn,superscriptaddress,showpacs]{revtex4}

\usepackage{dcolumn}

\usepackage{bm}

\usepackage{color}

\usepackage[english]{babel}

\usepackage{amsmath}

\usepackage{graphicx}

\usepackage{amssymb}

\usepackage{amsfonts}

\usepackage{subfigure}

\newcommand{\ket}[1]{\vert #1 \rangle}

\newcommand{\bra}[1]{\langle #1 \vert}

\newcommand{\ketbra}[2]{\vert #1 \rangle \! \langle #2 \vert}

\newcommand{\media}[1]{\langle #1 \rangle}

\newcommand{\Tr}[1]{\textrm{Tr}\,\! #1}

\newcommand{\beq}{\begin{equation}}

\newcommand{\eeq}{\end{equation}}

\newcommand{\barr}{\begin{eqnarray}}

\newcommand{\earr}{\end{eqnarray}}

\newcommand{\andy}[1]{ }

\def\bra#1{\langle #1 |}
\def\ket#1{| #1 \rangle}

\newcommand{\tr}{\mathop{\text{Tr}}\nolimits}

\begin{document}

\title{Entanglement of two blocks of spins in the critical Ising model}

\author{P. Facchi}
\affiliation{Dipartimento di Matematica, Universit\`a di Bari,
        I-70125  Bari, Italy}
\affiliation{INFN, Sezione di Bari, I-70126 Bari, Italy}
\author{G. Florio} \affiliation{Dipartimento di Fisica,
Universit\`a di Bari,
        I-70126  Bari, Italy}
\affiliation{INFN, Sezione di Bari, I-70126 Bari, Italy}
\author{C. Invernizzi}
\affiliation{Dipartimento di Fisica,
Universit\`a di Milano,
        I-20133 Milano, Italy }
\author{S. Pascazio} \affiliation{Dipartimento di Fisica,
Universit\`a di Bari,
       I-70126  Bari, Italy}
\affiliation{INFN, Sezione di Bari, I-70126 Bari, Italy}


\begin{abstract}
We compute the entropy of entanglement of two blocks of $L$ spins at
a distance $d$ in the ground state of an Ising chain in an external
transverse magnetic field. We numerically study the von
Neumann entropy for different values of the transverse field. At the
critical point we obtain analytical results for blocks of size $L=1$
and $L=2$. In the general case, the critical entropy is shown to be
additive when $d\rightarrow\infty$. Finally, based on simple
arguments, we derive an expression for the entropy at the critical
point as a function of both $L$ and $d$. This formula is in
excellent agreement with numerical results.
\end{abstract}


\maketitle

\section{Introduction}
\label{sec:Introduction}

A comprehension of the features of entanglement in systems with many
degrees of freedom, such as quantum spin chains, is currently one of
the most challenging problems, at the borderline of quantum
information science \cite{nielsen} and statistical physics. In the
last few years several measures of entanglement have been proposed
\cite{Osterloh, Osborne, LRV} and calculated (analytically in the
simplest cases, otherwise numerically) for the ground states of
many-body systems \cite{sarorev}.

Despite accurate investigations and different proposals, there is
still no consensus on the correct characterization of the
\emph{multipartite} entanglement of the ground state of a
many-body system. We will consider here the entanglement entropy,
a measure that can sometimes be tackled by analytic investigations
and for which quantum field theoretical methods can be employed.
The entanglement entropy is just the von Neumann entropy
associated with the reduced density matrix, that is the entropy of
a subsystem of the chain, and was explicitly evaluated for quantum
spin chains \cite{VLRK, LRV, Cardy&Calabrese, Korepin, eisert,
Its}.

One of the most striking features of the entanglement entropy is its
universal behavior at and close to a quantum phase transition.
Indeed, it is found that entropy in non-critical systems generally
tends to saturate towards a finite value as the size of the
subsystem increases, but this value (logarithmically) diverges with
the size of the subsystem as the system approaches a quantum
critical point. Close to a quantum critical point, where the
correlation length $\xi$ is much larger than the lattice spacing,
correlations are described by a $1+1$ dimensional quantum field
theory and at the critical point, where $\xi$ diverges, the field
theory is also a \textit{conformal} field theory \cite{cft}. In the
latter case, the behavior of entropy calculated by analytical and
numerical techniques for several spin systems is confirmed by the
predictions of the corresponding field theory.

In this work we extend the characterization of the entanglement
entropy to a more general subsystem, in which the correlations
between two disjoint blocks of spins and the rest of the chain are
studied as a function of the distance $d$ between the blocks and
their common size $L$. This entropy of entanglement will be denoted
$S(L,d)$ and will be analyzed by analytical and numerical methods.

The physical system we shall consider is the Ising model in a
transverse magnetic field, since it fulfills a convenient
combination of requirements. It is solvable, its ground state can
be computed by using well-known analytical and numerical
techniques \cite{VLRK} and, at the same time, it successfully
describes a rich spectrum of physical phenomena, that include the
ordered and disordered magnetic phases, connected by a quantum
phase transition \cite{sachdev}.

We will analytically compute the entropy of entanglement at the
quantum critical point (QCP) for blocks of $L=1$ and $L=2$ spins,
and will tackle the problem numerically for larger values of $L$.
We will first study the behavior of the entropy as a function of
the magnetic field $\lambda$, then at the QCP, $\lambda=1$, as a
function of the distance $d$ between the blocks and their size
$L$. We will investigate the limits $d\rightarrow 0$ and
$d\rightarrow\infty$. Our results will include as a particular
case ($d=0$) the logarithmic behaviour of the entropy of a single
block of $L$ spins at criticality $S_L =\frac{1}{6}\log L +
\mathcal{K}$, where $\mathcal{K}$ is a constant
\cite{LRV,VLRK,Cardy&Calabrese,eisert,Its,Korepin,Callan,FWilczek,Casini}.
We will also show the additivity of $S(L,d)$ at the critical point
as $d\rightarrow\infty$. Finally, we will plot $S(L,d)$ as a
function of both $L$ and $d$, getting an accurate idea of the
features of the entropy at the QCP.

This paper is divided in six sections. In Section \ref{ising} we
review previous works on spin chains, following
\cite{Lieb,Katsura,Barouch1, Barouch,VLRK,LRV}, where the
ground state of the Ising model is computed: the explicit
expressions obtained will be used in the following sections
in order to obtain the
reduced density matrix $\varrho_L$ and the Von Neumann entropy $S_L$
of $L$ contiguous spins. In
Section \ref{SdueBlocchi} we extend the definition of the
correlation matrix given in \cite{VLRK, LRV} to a bipartition of two
blocks of $L$ spins separated by a generic distance $d$.  We define
here the entropy $S(L,d)$, describing the entanglement of the two
blocks with the rest of the chain. By making use of the newly
defined reduced density matrix $\varrho_{L,d}$, in Section
\ref{analyticalRes}, we analytically compute the entropy for
blocks of one and two spins at the critical point. In Section
\ref{numericalRes} we carry out numerical computations of $S(L,d)$
for several sizes of the blocks. Finally, we plot the entropy as a
function of the size $L$ of the blocks and their reciprocal distance
$d$, in order to get a general idea of the features of the entropy
of entanglement at the critical point. Our results are summarized
and discussed in Section \ref{concl}. In the Appendices we included,
for self consistency, additional material and explicit calculations.


\section{Ground state of the Ising model}\label{ising}

The Ising chain in a tranverse field consists of $2 N +1$ spins with
nearest neighbor interactions and an external magnetic field,
described by the Hamiltonian
\begin{equation}\label{Hising}
\mathcal{H}_{I} = -J\sum_{-N\leq i \leq N} (\lambda \sigma_i^z  + \sigma_i^x \sigma_{i+1}^x).
\end{equation}
Here $i$ labels the  spins (we
take an odd number of spins for simplicity), $J>0$ and we consider open boundary
conditions, $\sigma_{N+1}^x=0$. $\sigma_{i}^{\mu}$ ($\mu=x,y,z$) are the Pauli
matrices acting on spin $i$. The determination of the ground state
proceeds with the Jordan-Wigner transformation in terms of Dirac
or Majorana fermionic operators
\cite{Lieb,Katsura,Barouch1,Barouch}. Here it will be convenient to consider Majorana fermions, whose operators are defined by
\begin{equation}
\label{Majorana}
\check{a}_{2l-1}\equiv \left( \prod_{m<l} \sigma_m ^z \right)\sigma_l ^x ,
\qquad \check{a}_{2l}\equiv\left(\prod_{m<l}\sigma_m ^z \right)\sigma_l ^y ,
\end{equation}
with $-N\leq l\leq N$.
They are hermitian and obey anticommutation relations,
\begin{equation}
\check{a}_m^\dagger=\check{a}_m, \qquad
\{\check{a}_m,\check{a}_n\}=2\delta_{mn},
\end{equation}
 and their expectation values in the ground state $\ket{\psi_0}$
\begin{equation}\label{correlationfunction}
\bra{\psi_0}\check{a}_m\check{a}_n\ket{\psi_0}=\media{\check{a}_m\check{a}_n}=\delta_{mn} + i
\Gamma^A_{mn},
\end{equation}
with $-2 N -1 \leq m, n \leq 2 N$, completely characterize $\ket{\psi_0}$.
Consider now a block  of $L$ contiguos spins labeled by $i$ with
\begin{equation}
 k\leq i \leq k + L - 1,
\label{eq:blockrange}
\end{equation}
with $k>-N$ and $k + L - 1<N$. The expectation values of the Majorana operators of the block are encoded in the $2L\times 2L$ submatrix
\begin{equation}
\left(\Gamma^A_L\right)_{mn} = -i (\media{\check{a}_m\check{a}_n}-\delta_{mn}),
\end{equation}
with $2 k -1 \leq m, n \leq 2 k + 2 L -2$.

We are interested in the thermodynamic limit of an infinite chain,
$N\to\infty$. In such a limit, the ground state becomes translation
invariant,  and all correlations inherit such an invariance:
$\media{\check{a}_{2m} \check{a}_{2n}}=\media{\check{a}_{2m-1}
\check{a}_{2n-1}} = 0$, $\forall m, n$ with $m\neq n$, while
$\media{\check{a}_{2 m-1} \check{a}_{2n}}= i g_{m-n}$ depend only on
the difference $m-n$. Therefore, the block correlation matrix
$\Gamma^A_L$ becomes independent of $k$ and reads
\begin{equation}\label{Gamma_A}
\Gamma^A_L= \left (\begin{array}{cccc} \Pi_0 & \Pi_{-1} & \ldots & \Pi_{-L+1} \\
\Pi_1 & \Pi_0 &  & \vdots \\
\vdots & & \ddots & \vdots \\
\Pi_{L-1} & \ldots & \ldots & \Pi_0 \end{array} \right ) ,
\end{equation}
with
\begin{equation}
\Pi_l=
\left (\begin{array}{cc}
0 & -i \media{\check{a}_{2l-1}\check{a}_0} \\  -i \media{\check{a}_{2l}\check{a}_{-1}}& 0 \end{array} \right )
=\left (\begin{array}{cc}
0 & g_l \\ -g_{-l} & 0 \end{array} \right ), \label{Gamma_AB}
\end{equation}
where
the real coefficients $g_l$ are given, for an infinite
chain, by
\begin{equation}\label{g_l}
g_l= \frac{1}{2\pi}\int_{0} ^{2\pi} d\phi \; e^{i \phi l}
\frac{(\cos{ \phi} -\lambda)  + i \sin {\phi}}{\sqrt{(\cos\phi - \lambda )^2 + \sin^2{\phi}}}.
\end{equation}
Thus, $\Gamma^A_L$ is a real, skew-symmetic $2L\times
2L$ matrix, i.e. $(\Gamma^A_L) ^T = -\Gamma^A_L$, since all the
blocks $\Pi_l$ have the property $(\Pi_l )^T= - \Pi_{-l} $.

\begin{figure}
\includegraphics[width=0.5 \textwidth]{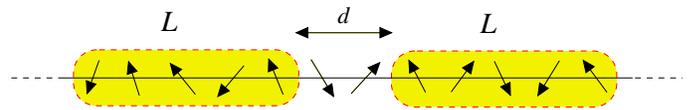}
\caption{(Color online) Two blocks of $L$ adjacent spins at a distance
$d$. The state $\varrho_{L,d}$ is obtained from the ground state
$\ket{\psi_0}$ of the
spin chain by tracing out the
spins that do not belong to the blocks.}\label{2blocchi}
\end{figure}

\section{Entropy of Two Blocks of Spins}\label{SdueBlocchi}

The entropy of a single block of $L$ contiguous spins in the
critical regime can be obtained by very accurate numerical results
\cite{LRV, VLRK} and analytical conformal field theory
calculations \cite{Cardy&Calabrese, Korepin}. Given the ground state
$\ket{\psi_0}$, one finds the reduced density matrix
\begin{equation}
\varrho_L=\tr_{\neg L}\{\ket{\psi_0}\bra{\psi_0}\},
\end{equation}
where the trace is over all spins that do not belong to the block,
and its entropy
\begin{eqnarray}
\label{S_Wilczek}
S_L =-\tr\{\varrho_L\log\varrho_L\} .
\end{eqnarray}
For definiteness, in this paper we will fix the base of logarithms
to 2. The calculations of Refs.\
\cite{VLRK,LRV}, yielding an expression of the reduced density
matrix $\varrho_L$ and the entanglement entropy $S_L$ of $L$
adjacent spins in the ground state $\ket{\psi_0}$ are reviewed in
Appendix \ref{computationS}. The key point is the following: the
block entropy (\ref{S_Wilczek}) is the sum of $L$ terms
\begin{equation}\label{def:S_L}
S_L =\sum_{l=1} ^L H\left(\frac{1+\nu_l}{2}\right),
\end{equation}
where
\begin{equation}
\label{eq:Shannon}
H(x)= - x \log x- (1-x) \log (1-x)
\end{equation}
is the  Shannon entropy of a bit,
and $\pm i \nu_l$,  with $1\leq l\leq L$, are the pairs of (purely imaginary)
eigenvalues of the block correlation matrix $\Gamma^A_L$ of Eq.\ (\ref{Gamma_A}).

In the continuous limit
\begin{eqnarray}
\label{S_Wilczekaaa}
S_L =\frac{1}{6}\log L +
\mathcal{K}(L),
\end{eqnarray}
where $1/6$ derives from the central charge $c=1/2$ of a free
massless fermionic field and
\begin{eqnarray}
\label{KL}
\mathcal{K}(L)=
\mathcal{K}+O\left(\frac{1}{L}\right), \qquad L\to\infty,
\end{eqnarray}
$\mathcal{K}$ being a constant.
In this section we want to extend this approach and construct the
density matrix $\varrho_{L,d}$ of a subsystem of two blocks of $L$
adjacent spins situated at a distance $d$, studying their
entanglement with the rest of the chain. See Fig.\ \ref{2blocchi}.
To this aim, one starts by computing the  matrix $\Gamma_L^A$
of a single block of adjacent spins
(\ref{Gamma_A}), and then traces out the central $d$ spins as
follows.

We define the $4L \times 4L$ correlation matrix $\Gamma_{L,d} ^A$
of two blocks, each of $L$
spins, situated at a distance $d$ ($d$, like $L$, are
expressed in units of the distance between adjacent spins,
and are therefore dimensionless)
\begin{eqnarray}\label{GammaLd}
\Gamma_{L,d} ^A  =  \left (\begin{array}{cc}
A_0^{(L)} & A_{-L-d}^{(L)} \\
 & \\
A_{L+d}^{(L)} & A_0^{(L)}
\end{array}\right )
=  \left (\begin{array}{cc}
\Gamma^A_L & A_{-L-d}^{(L)} \\
 & \\
A_{L+d}^{(L)} & \Gamma^A_L
\end{array}\right ),
\end{eqnarray}
where $A_x^{(L)}=A_x^{(L,L)}$ with
\begin{eqnarray}
A_x^{(L,M)} =& \left (\begin{array}{cccc}
\Pi_x & \Pi_{x -1} & \ldots & \Pi_{x - M + 1} \\
\Pi_{x+1} & \Pi_x & \ldots & \Pi_{x -M+2} \\
\vdots &    & \ddots &  \vdots   \\
\Pi_{x+L-1} & \Pi_{x+L-2} & \ldots & \Pi_{x-M+L}
\end{array}\right ) .
\label{eq:AxLM}
\end{eqnarray}
The matrix $A_x^{(L)}= \left (- A_{-x}^{(L)}\right )^T$ is a Toeplitz matrix, and
$\Gamma_{L,d} ^A$ has the property
\begin{equation}\label{Gamma(L,0)}
\Gamma_{L,0}^A = \Gamma_{2L} ^A,
\end{equation}
i.e., when the distance between the two blocks is zero,
$\Gamma_{L,d} ^A$ becomes equal to the matrix
(\ref{Gamma_A})
of a single block of size $2L$. The matrix $\Gamma_{L,d}^A$ in Eq.\
(\ref{GammaLd}) is obtained by tracing out the $d$ rows and $d$
columns that are labeled with $L \leq x < L+d$ in the $(2L +d)
\times (2L+d)$ matrix $\Gamma_{2L + d} ^A$,
\begin{eqnarray}\label{Gamma2Lpiud}
\Gamma_{2L+d} ^A =  \left (\begin{array}{ccc}
A_0^{(L)} &  A_{-L}^{(L,d)} & A_{-L-d}^{(L)} \\
& & \\
A_{L}^{(d,L)}    &  A_{0}^{(d)}  &  A_{-d}^{(d,L)}   \\
& & \\
A_{L + d}^{(L)} & A_{d}^{(L,d)}  & A_0^{(L)}
\end{array}\right ).
\end{eqnarray}
For example, let us consider the case of two blocks of $L = 2$ spins
at a distance $d=3$. In this case the $7\times 7$ matrix
$\Gamma_{2L + d} ^A=\Gamma_7 ^A$ reads
\begin{eqnarray}
\Gamma_7 ^A = \left (\begin{array}{ccccccc} \Pi_0 & \Pi_{-1} & \Pi_{-2} &\Pi_{-3}& \Pi_{-4} & \Pi_{-5} & \Pi_{-6} \\
\Pi_1 & \Pi_0 &  \Pi_{-1}&\Pi_{-2}&\ldots &  & \Pi_{-5} \\
\Pi_{2} & \Pi_{1} & \ddots & & & & \vdots \\
\Pi_3 & \Pi_{2} &  &\ddots & & &\vdots \\
\Pi_{4} &  \vdots & & & & &\vdots \\
\Pi_{5} & & & & & \ddots & \vdots\\
\Pi_{6} &   \ldots &  &\ldots & & \ldots & \Pi_0 \end{array} \right )
\end{eqnarray}
and we have to cancel the columns whose first element is labelled by
$-2$, $-3$, $-4$ and the rows whose first element is labelled by
$2$, $3$, $4$, obtaining
\begin{eqnarray}
\Gamma_{2,3} ^A = \left (\begin{array}{cccc} \Pi_0 & \Pi_{-1} &\Pi_{-5} & \Pi_{-6} \\
\Pi_1 & \Pi_0 & \Pi_{-4} & \Pi_{-5} \\
\Pi_5 & \Pi_4 & \Pi_0 & \Pi_{-1} \\
\Pi_{6} & \Pi_5 & \Pi_1 & \Pi_0 \end{array} \right ),
\end{eqnarray}
that is again a real skew-symmetrix matrix.

The entanglement of the two blocks of spins reads
\begin{equation}\label{S(L,d)}
S(L,d)=-\Tr(\varrho_{L,d}\log\varrho_{L,d}),
\end{equation}
where $\varrho_{L,d}$ is the density matrix of two blocks of $L$
adjacent spins at a distance $d$. Exactly as for a single block, the above entropy can be given an
explicit expression in terms of the eigenvalues $\pm i \nu_l$, with $1\leq l\leq 2L$, of
$\Gamma_{L,d} ^A$, analogous to (\ref{def:S_L}),
\begin{equation}\label{def:S_Ld}
S(L,d) =\sum_{l=1} ^{2L} H\left(\frac{1+\nu_l}{2}\right),
\end{equation}
with $H$ given by (\ref{eq:Shannon}).

Before investigating the behavior of Eq.\ (\ref{S(L,d)}),
it is instructive to look first at some simple examples.

\section{Analytical Results}\label{analyticalRes}

\subsection{Entanglement of Two Single Spins }

We consider the Ising chain in a critical transverse magnetic field
$\lambda_c=1$. At $\lambda = \lambda_c$, the coefficients $g_l$ of the
reduced correlation matrix $\Gamma_{L,d} ^A$, defined in Section
\ref{SdueBlocchi}, and given by Eq.\ (\ref{g_l}),
can be computed analytically, yielding
\begin{equation}
\label{gl}
g_l =
\frac{1}{2\pi}\int_{0} ^{2\pi} d\phi \; e^{i \phi l} i e^{i \frac{\phi}{2}}= -\frac{1}{\pi (l + \frac{1}{2})}.
\end{equation}

In order to compute the entanglement entropy $S(L,d)$ of two single
spins at a distance $d$, we have to calculate the eigenvalues of the
correlation matrix $\Gamma_{L,d} ^A $, where $L = 1$ and $d\geq0$.
Equation (\ref{GammaLd}) reads
\begin{eqnarray}
\Gamma_{1,d} ^A =& \left ( \begin{array}{cc}
\Pi_0 & \Pi_{-(1+ d)} \\
\Pi_{1 + d} & \Pi_0 \end{array} \right ) \nonumber \\
=& \left ( \begin{array}{cccc}
0 & g_0 & 0 & g_{-l} \\
-g_0 & 0 & -g_l & 0 \\
0 & g_l & 0 & g_0 \\
-g_{-l} & 0 & -g_0 & 0
\end{array} \right ) \nonumber \\
=& \left ( \begin{array}{cccc}
0 & \frac{-2}{\pi} & 0 & \frac{-2}{\pi(-2 l + 1)} \\
\frac{2}{\pi} & 0 &\frac{2}{\pi(2 l + 1)} & 0 \\
0 & \frac{-2}{\pi(2 l + 1)} & 0 &\frac{-2}{\pi}  \\
\frac{2}{\pi (-2 l + 1)} & 0 & \frac{2}{\pi} & 0
\end{array} \right ),
\end{eqnarray}
where $l = d+1$.
The eigenvalues are the solutions to the characteristic equation
\begin{equation}\label{eq_caratteristica}
\det (\Gamma_{L,d} ^A - \mu ) = 0.
\end{equation}
For $L=1$, there are four eigenvalues $\pm i \nu_1$ and  $\pm i\nu_2$, with
\begin{equation}
 \nu_{1,2} =
\frac{2}{\pi}\frac{\sqrt{(4 l^2-1)^2+4l^2}\pm 1}{4 l^2-1}.
 \label{eq:eigenvalues1+1}
\end{equation}
The eigenvalues of the reduced density matrix $\varrho_{L,d}$ are $(1 \pm  \nu_k)/2$
and the von Neumann entropy reads
\begin{equation}\label{entropy1spin}
 S(1,d) =  H\left(\frac{1+\nu_1}{2}\right)+H\left(\frac{1+\nu_2}{2}\right),
 \end{equation}
 where $H$ is given by (\ref{eq:Shannon}).

It is interesting to consider the cases $d=0$ and $d\gg 1$. In the
former case we have
 $\nu_{1,2} =2(\sqrt{13}\pm 1)/(3\pi)$, whence
\barr
S(1,0) &=& H\left(\frac{\sqrt{13}+2}{3\pi}\right)+H\left(\frac{\sqrt{13}}{3\pi}\right)
\nonumber\\
& =&-\frac{1}{2} \log \left(\frac{1}{16}+\frac{16-7 \pi ^2}{9 \pi ^4}\right)\nonumber\\
   & & - \frac{\sqrt{13} }{3 \pi  }\log \left(1+\frac{8
    \sqrt{13} \pi }{16-4 \sqrt{13} \pi +3 \pi ^2}\right) \nonumber\\
   & & - \frac{1}{3 \pi }\log \left(1+\frac{8 \pi }{3 \pi^2  -4
   \pi-16}\right).
   \earr
In the latter case, since the eigenvalues $\nu_{1,2}=2/\pi + O(1/d^2)$ coincide in the limit
$d\rightarrow+\infty$,
we have
\barr
S(1,d) &=& 2 H\left(\frac{\pi+2}{2\pi}\right) +O\left({\frac{1}{d^2}}\right)\nonumber\\
&=&
\frac{2}{\pi}\log\left(\frac{\pi-2}{\pi+2}\right)+\log\left(\frac{4\pi^2}{\pi^2-4}\right)+O\left({\frac{1}{d^2}}\right).
\nonumber\\
\earr
 More on this phenomenon later.

\subsection{Entanglement of two blocks of $L=2$ spins}
\label{Two_Blocks}

If $L=2$, we have to compute the eigenvalues of
\begin{eqnarray}
\Gamma_{2,d} ^A =& \left ( \begin{array}{cccc}
\Pi_0 & \Pi_{-1} & \Pi_{-(2 + d)} & \Pi_{-(3+d)} \\
\Pi_1 & \Pi_0 & \Pi_{-(1 + d)} & \Pi_{- (2+d)} \\
\Pi_{(2+d)} & \Pi_{(1 + d)} & \Pi_0 & \Pi_{-1} \\
\Pi_{(3+d)} & \Pi_{(2 +d)} & \Pi_{1 } & \Pi_0
\end{array} \right ) \nonumber \\
=& \left ( \begin{array}{cccc}
\Pi_0 & \Pi_{-1} & \Pi_{-(1 + l)} & \Pi_{-(2+l)} \\
\Pi_1 & \Pi_0 & \Pi_{- l} & \Pi_{- (1+l)} \\
\Pi_{(1+l)} & \Pi_{l} & \Pi_0 & \Pi_{-1} \\
\Pi_{(2+l)} & \Pi_{(1+l)} & \Pi_{1 } & \Pi_0
\end{array} \right ),
\end{eqnarray}
with $l=d+1$.
The characteristic equation (\ref{eq_caratteristica}) is of $8$th degree, but can be reduced to a quartic equation in $t = \mu^2$
\begin{equation}\label{quartic}
t^4 + p t^3 + q t^2 + r t + s = 0,
\end{equation}
which has an exact solution. The coefficients $p$, $q$, $r$ and $s$
are functions of the distance  $d$ and are explicitly written in
Appendix \ref{eigenvalues}.

The matrix $\Gamma_{L,d} ^A$ will have the eight eigenvalues $\pm \mu_k
= \pm i\nu_k$, with $k=1,2,3,4$, two for each (negative) root of $t$. The
eigenvalues of the reduced density matrix $\varrho_{2,d}$ are
$(1\pm \nu_k)/2$ and the entropy reads
\begin{equation}
S(L,d) =  \sum_{k=1} ^4 H\left(\frac{1+\nu_k}{2}\right),
\end{equation}
with $H$ given by (\ref{eq:Shannon}).

\begin{figure}
\includegraphics[width=0.48\textwidth]{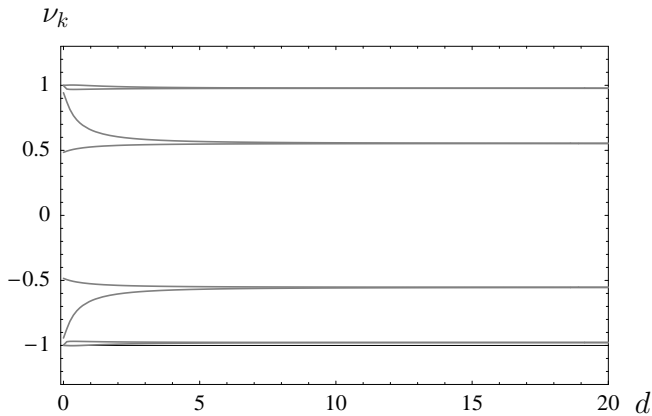}
\caption{(Color online) Eigenvalues of the reduced density matrix
of two blocks of $L=2$ spins versus their distance $d$ at the
critical point $\lambda_c=1$. Note that the $\nu_k$'s reach a
saturation value at $d \simeq 5$.
Here and in the following figures, $d$
is expressed in units of the distance between adjacent spins,
and is therefore dimensionless.} \label{autovL2}
\end{figure}

In Fig.\ \ref{autovL2} we plot the eigenvalues $\nu_k$ versus the
distance $d$. Note that the eigenvalues quickly saturate at a
distance $d \simeq 5$ between the blocks. This means that the
entanglement between the two $L=2$ blocks reaches its asymptotic
value for $d \gtrsim 5$.

The asymptotic values of the eigenvalues are solutions to the equation
obtained by taking the limit $d\to\infty$ of (\ref{quartic}). One gets
\begin{eqnarray}
\nu_{1}(\infty)&=&\nu_{3}(\infty)=\frac{2}{\pi} \frac{\sqrt{13}+1}{3},
\nonumber\\
\nu_{2}(\infty)&=&\nu_{4}(\infty)=\frac{2}{\pi} \frac{\sqrt{13}-1}{3},
\end{eqnarray}
in agreement with Fig.\ \ref{autovL2}. Note that they coincide with
(\ref{eq:eigenvalues1+1}) evaluated at $l=1$, i.e.\ with the
eigenvalues of the reduced density matrix of two spins at a distance
$d=l-1=0$. Therefore, for $d\to \infty$  the eigenvalues coalesce
into pairs and the spectrum of two blocks (of two spins) coincides
with the spectrum (with degeneracy 2) of a single block. This
phenomenon, which implies the asymptotic additivity of block
entropy, is independent of the blocks dimension $L$ and will be
discussed in full generality in Sec.\ \ref{sec:additivity}.

\section{Entropy of two blocks of spins}\label{numericalRes}

In this section we look at the entanglement entropy $S(L,d)$ of two
blocks of spins when the magnetic field varies.
At the critical point,
$\lambda_c=1$, we will find an expression for $S(L,d)$ in terms of
the entropy $S_L$ of a single block, given in Eqs.\
(\ref{S_Wilczek})-(\ref{S_Wilczekaaa}), and investigate its limits
$d\rightarrow 0$, $d\rightarrow \infty$. We will combine numerical
estimates with analytical methods.

\subsection{Entropy versus $\lambda$}
\label{entlambda}

We start by evaluating the entanglement entropy $S(L,d)$ versus the
magnetic field $\lambda$. The entanglement between two single spins
at a distance $d$ and the remaining part of the chain was
investigated in Refs.\
\cite{RigolinPRA, RigolinPRArapid, RigolinPRL} as a function
of the magnetic field $\lambda$. A generalization to a comb of $m$ spins, spaced $d$ sites apart can be found in \cite{Keating}. We now generalize these results to
the case of two arbitrary blocks of $L$ spins at a distance $d$.

\begin{figure}
\includegraphics[width=.5\textwidth]{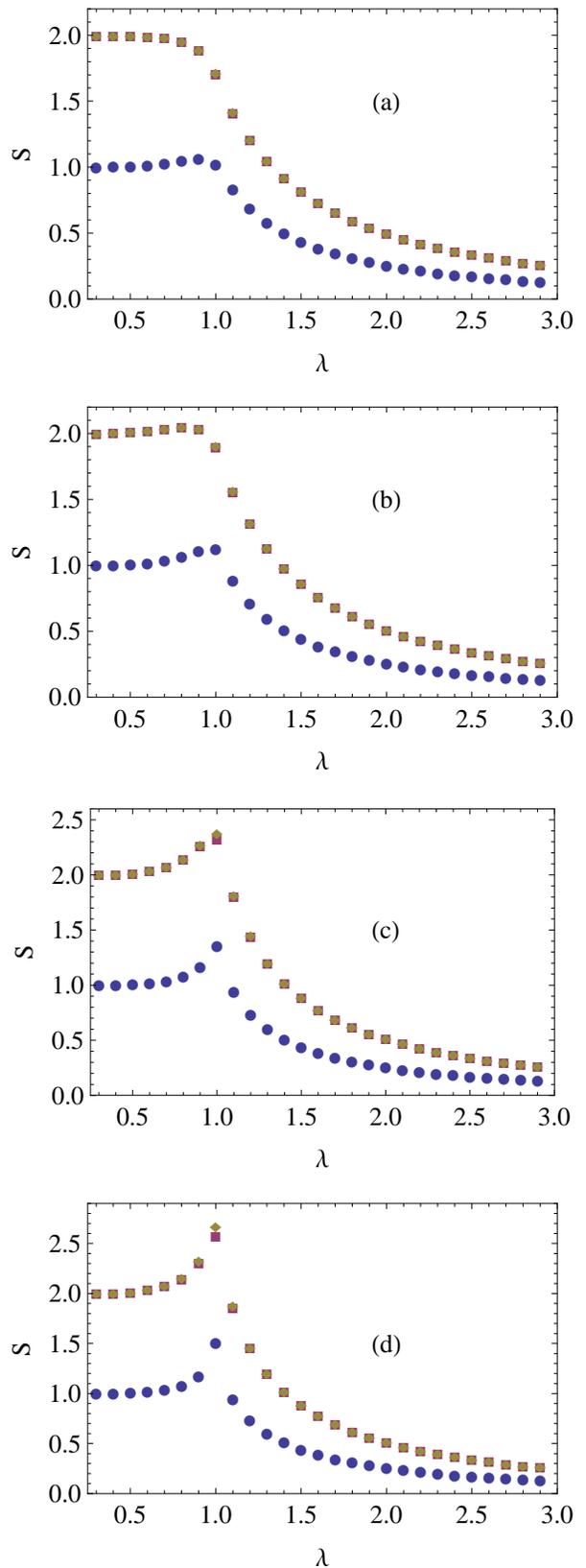}
\caption{(Color online) Entropy of the reduced density matrix for
blocks of (a) $L=2$, (b) $L=3$, (c) $L=8$ and (d) $L=15$ spins versus the external magnetic field
$\lambda$ for several distances $d$ between the blocks.
Circles $d=0$, squares $d=10$, diamonds $d=50$.
Squares and diamonds are indistinguishable in (a) and (b), and are barely distinguishable
in (c) and (d) only around the critical value $\lambda=1$.}\label{S:lambda}
\end{figure}

In general, the presence of a gap between the blocks yields
a larger entropy for all values of the magnetic field $\lambda$. Let
us start examining  the situation at zero magnetic field. At
$\lambda=0$, from (\ref{g_l}) one gets  that $g_l= \delta_{l,-1}$.
Therefore, since $\Pi_l=0$ for $l\neq\pm 1$ and
$\Pi_{\pm 1}=\sigma^{\mp}=(\sigma_1 \mp i\sigma_2)/2$,
Eqs.\
(\ref{GammaLd}) and (\ref{eq:AxLM}) greatly simplify. In particular,
for any $L>1$, the off-diagonal blocks in (\ref{GammaLd}) read, for
$d=0$,
\begin{equation}
A_{L}^{(L)} = - \left(A_{-L}^{(L)}\right)^T =\left (\begin{array}{cccc}
0 & 0 & \ldots & \Pi_{+1} \\
\vdots &    & \ddots &  \vdots   \\
0 &0 & \ldots & 0
\end{array}\right )
\end{equation}
and yield a tridiagonal block matrix
\begin{equation}
\Gamma^{A}_{L,0} = \Gamma^A_{2L} =
\left (\begin{array}{ccccc}
0 & \Pi_{-1} & 0 & \ldots & 0 \\
\Pi_{+1} & 0 &  \Pi_{-1} & \ldots & 0 \\
\vdots & \vdots &     \ddots & &  \vdots   \\
0 &0 & 0 & \ldots & 0
\end{array}\right ) ,
\end{equation}
while, for any $d>0$,
$A_{L+d}^{(L)} = - \left(A_{-L-d}^{(L)}\right)^T=0$ and one gets
\begin{equation}
\Gamma^{A}_{L,d}= \Gamma^{A}_{L} \oplus \Gamma^{A}_{L}.
\end{equation}
Thus, at $d=0$, the characteristic polynomial is
\begin{equation}
\det (\Gamma_{L,0} ^A - \mu ) =\det (\Gamma_{2L} ^A - \mu )= \mu^2 (\mu^2 +1)^{2L-1} ,
\end{equation}
hence $\nu_1=0$ and $\nu_l=1$ for $2\leq l\leq 2 L$, so that
\begin{equation}
S(L,0)=H\left(\frac{1}{2}\right)+\sum_{l=2}^{2 L} H(1)=H\left(\frac{1}{2}\right)=\log 2=1.
\end{equation}
On the other hand, for $d\geq 1$, one gets
\begin{equation}
\det (\Gamma_{L,d} ^A - \mu ) =\det (\Gamma_{L} ^A - \mu )^2= \mu^4 (\mu^2 +1)^{2L-2} ,
\end{equation}
hence $\nu_1=\nu_2 = 0$ and $\nu_l=1$ for $3\leq l\leq 2 L$ and
\begin{equation}
S(L,d)= 2 S(L,0) =
2 H\left(\frac{1}{2}\right)=2.
\end{equation}
This is intuitively clear: at zero transverse field $S(L,d)$ follows
exactly an area law \cite{arealawrev} and in 1 dimension the presence of a gap doubles
the area of the boundary, doubling the entropy. See Fig.\
\ref{S:lambda}. For nonzero values of the magnetic field, there are
correction to the area law, due to correlations between the two
blocks. An entropy increase is still natural, but it turns out to be
smaller than the factor 2 that one would naively expect for a
doubled boundary. We will show that the factor 2 can be recovered in
the limit of large gap $d$: this is the phenomenon
of asymptotic additivity of entropy mentioned at the end of Sec.\ \ref{Two_Blocks}. See Sec.\
\ref{Twobcrit}.

We show the results of some numerical investigations
in Fig. \ref{S:lambda}. Let us first look at the case of small $L$ $(\simeq 2\div 3)$. For
$d>0$ entropy is not maximum at the critical point $\lambda=1$, but
rather for some $\lambda<1$. On the other hand, for bigger $L$, the
maximum entropy is always reached at the critical point $\lambda=1$
and its value grows with the size $L$ of the blocks.

One also notices that, far from criticality, the entropy has a very
weak dependence on the value of the gap $d\geq 1$, and thus an area
law is a very good approximation. The largest deviations are at the
critical point, when the correlation length diverges. At fixed $L$,
the ordinate of the cusp at $\lambda =1$ is always an increasing
function of $d$. We now turn to the study of the critical case and
endeavor to find some interesting analytical expressions.

\subsection{Critical chain}
\label{Twobcrit}

\subsubsection{Blocks of contiguous spins ($d\rightarrow0$)}
\label{bcs}

In the limit
$d\rightarrow 0$, the entropy of entanglement (\ref{S(L,d)}) must
reproduce the single-block result (\ref{S_Wilczek}) as a particular
case:
\begin{equation}
\label{fit1block}
S(L,0) = S_{2L} .
\end{equation}
This is a simple consistency check and was numerically verified when
the magnetic field is critical, $\lambda_c = 1$. As a byproduct,
this enables us to obtain the value of the constant $\mathcal{K}$
via the logarithmic fit
\begin{equation}\label{fit1block1}
S(L,0) = \frac{1}{6} \log 2L +
 \mathcal{K}+O\left(\frac{1}{L}\right).
\end{equation}
We obtain
\begin{eqnarray}
\mathcal{K} =0.690413,
\label{Knum}
\end{eqnarray}
with an error $\simeq 9 \cdot 10^{-6}$, corroborating the results in
\cite{LRV}. An accurate fit enables us to give a precise estimate of
the corrections in $1/L$, but more on this later.

\subsubsection{Asymptotic additivity of entropy ($d\rightarrow\infty$)}
\label{sec:additivity}
The plot in Fig.\ \ref{S(L,inf)num} shows that in the limit
$d\rightarrow\infty$ the two-block entropy is accurately fitted by
\begin{equation}
\label{fitSinf}
S(L,\infty) = \frac{1}{3}\log L + 2\mathcal{K}(L) = 2S_L.
\end{equation}
Entropy becomes therefore additive at large distances $d$. The
physical meaning of this result is that the entanglement entropy of
two separated blocks of $L$ spins becomes twice the entropy of a
single block $L$ when the distance between the blocks approaches
$\infty$, i.e. it becomes much larger than the size of a block.
Therefore, at the critical point, the quantum correlations between
two finite blocks of spins saturate at a certain distance. We
now explain this result, which turns out to be valid for every value
of the magnetic field $\lambda$.

Note that from (\ref{g_l}) one gets
\begin{equation}
\lim_{l\to\infty} g_l=0,
\end{equation} by Riemann-Lesbegue lemma. Thus,
from (\ref{Gamma_AB})
\begin{equation}
\lim_{l\to\infty} \Pi_l =0,
\end{equation}
and from (\ref{eq:AxLM})
\begin{equation}
\lim_{x\to\infty} A_x^{(L)}=0, \qquad \forall L.
\end{equation}
$A_x^{(L)}$ accounts for the residual correlation of two blocks
of spin at a distance $x$.

Therefore, the limit of the reduced correlation matrix of two blocks
(\ref{GammaLd}) reads
\begin{eqnarray}
\Gamma^A_{L,\infty} &=& \lim_{d\to+\infty} \Gamma^A_{L,d}= \lim_{d\to+\infty}
\left (\begin{array}{cc}
A_0^{(L)} & A_{-L-d}^{(L)} \\
 & \\
A_{L+d}^{(L)} & A_0^{(L)}
\end{array}\right )
\nonumber\\
&=&\left (\begin{array}{cc}
A_0^{(L)} & 0 \\
 & \\
0 & A_0^{(L)}
\end{array}\right ) =  \Gamma_L^A \oplus \Gamma_L^A,
\end{eqnarray}
for all $L$. The two blocks become independent and the limiting
spectrum of $\Gamma^A_{L,d}$ is given by the spectrum of a single
block $\Gamma_L^A$, with degeneracy 2. As a consequence, the entropy
(\ref{def:S_Ld}) becomes additive in the limit
\begin{eqnarray}
S(L,\infty) &=&\sum_{l=1} ^{2L} H\left(\frac{1+\nu_l(\Gamma^A_{L,\infty})}{2}\right)
\nonumber\\
&=& 2 \sum_{l=1} ^{L} H\left(\frac{1+\nu_l(\Gamma^A_L)}{2}\right)= 2 S_L,
\end{eqnarray}
where
$\pm i \nu_l(\Gamma^A_{L,d})$ with $1\leq l\leq 2L$ denote the eigenvalues of $\Gamma^A_{L,d}$
and $\pm i \nu_m(\Gamma^A_{L})$ with $1\leq m\leq L$ denote the eigenvalues of $\Gamma^A_{L}$.
We stress again that these results are valid for all values of the magnetic
field $\lambda$. See the introductory comments in Sec.\ \ref{entlambda}. The area law discussed there is restored for sufficiently large $d$ when the correlations between the two blocks are negligible and the two boundaries become ``independent".


\begin{figure}
\includegraphics[width=0.48\textwidth]{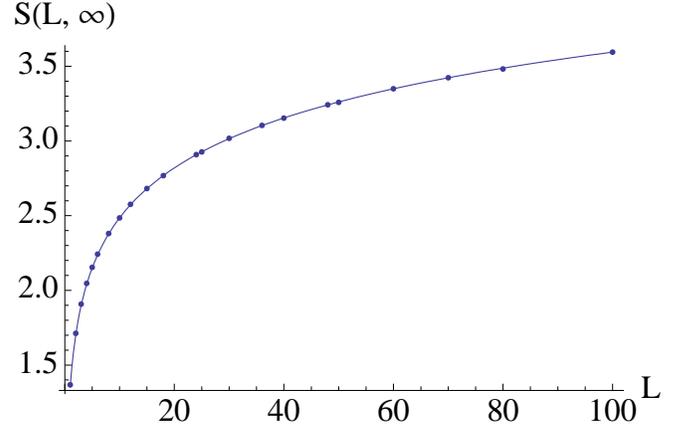}
\caption{(Color online) Saturation value of the critical block entropy ($\lambda=1$).
The numerical values of $S(L,\infty)$ are fitted by $\frac{1}{3}\log
L + 2\mathcal{K}(L) $.}
\label{S(L,inf)num}
\end{figure}

\subsubsection{General behavior of $S(L,d)$ at the critical point}

We now turn to the problem of describing the entanglement of two
blocks of $L$ spin at an arbitrary distance $d$ with the remaining
part of the critical Ising chain. In order to find a function of $L$
and $d$ we make two assumptions: we require that the entropy be a
function of \textit{all} the scales of the problem; moreover, the
dependence must be logarithmic.

Let therefore
\begin{eqnarray}
S(L,d)&=&\frac{1}{6}\Big(2\log(L-\alpha)-2\log(L+d) \nonumber\\
& & + \log(2L+d-\alpha) +\log(d+\alpha) + \beta\Big), \quad
\label{eq:entropy}
\end{eqnarray}
with $0<\alpha<1$ and $\beta\in\mathbb{R}$. See Fig.\ \ref{2blocchi}. The quantity $\alpha$
fixes the position of the end of each block, $\alpha=0$
corresponding to the central point between two adjacent spins, while
$\alpha=1$ to the position of the last (or first) spin. Clearly,
$\alpha$ detects granularity in the chain and the corrections due to
$\alpha$ will be important for small values of $L$ and/or $d$. We
obtain
\begin{eqnarray}
S(L,0)&=&\frac{1}{6}\log(2L-\alpha)+\frac{1}{3}\log\left(1-\frac{\alpha}{L}\right)
\nonumber\\
& &+\frac{1}{6}(\log\alpha+\beta)
\nonumber\\
&\sim& \frac{1}{6}\log(2L)+\frac{1}{6}(\log\alpha+\beta), \quad L\to\infty,
\nonumber\\
\end{eqnarray}
whence
\begin{equation}
\log\alpha+\beta=6\mathcal{K} .
\end{equation}
On the other hand,
\begin{eqnarray}
S(L,+\infty)&=&\frac{1}{6}\Big(2\log(L-\alpha)+\beta\Big)
\nonumber\\
&\sim& \frac{1}{3}\log(L)+\frac{1}{6}\beta,  \quad L\to\infty,
\end{eqnarray}
whence
\begin{equation}
\beta=12\mathcal{K}.
\end{equation}
In conclusion,
\begin{eqnarray}
S(L,d)&=&\frac{1}{6}\Big(2\log(L-\alpha)-2\log(L+d) \nonumber\\
& & + \log(2L+d-\alpha) +\log(d+\alpha) -2\log\alpha \Big), \nonumber\\
\label{eq:entropy1}
\end{eqnarray}
with
\begin{equation}
\label{eq:alpha}
\alpha=2^{-6 \mathcal{K}}
= 0.0566226 .
\end{equation}
Notice that there are \emph{no} free parameters. Moreover, in the
realm of validity of CFT, when $d,L\gg 1$, one gets
\begin{eqnarray}
S(L,d)&\sim&\frac{1}{6}\Big(2\log L-2\log(L+d) \nonumber\\
& & + \log(2L+d) +\log d \Big) +\mathcal{K}, \nonumber\\
\label{eq:CC}
\end{eqnarray}
that agrees with the results of Calabrese and Cardy
\cite{Cardy&Calabrese} when one adds a missing
addendum in their formula (3.32).

From (\ref{eq:entropy1}) we get
\begin{equation}
S(L,0)=\frac{1}{6} \log(2L) +\mathcal{K}(L),
\end{equation}
where
\begin{eqnarray}
\mathcal{K}(L)&=& -\frac{1}{6}\log\alpha+\frac{1}{6}\log\left(1-\frac{\alpha}{2 L}\right)+
\frac{1}{3}\log\left(1-\frac{\alpha}{L}\right)
\nonumber\\
&\sim&\mathcal{K} -
\frac{5}{12}\frac{\alpha}{L}-\frac{3}{16}\frac{\alpha^2}{L^2}
-\frac{17}{144}\frac{\alpha^3}{L^3}, \quad L\to\infty.
\label{eq:KLexp}
\end{eqnarray}
This formula is in excellent agreement with numerical results.
It provides the explicit expression in Eq.\ (\ref{KL}) and was used
in Eqs.\ (\ref{fit1block1})-(\ref{Knum}) and Fig.\
\ref{S(L,inf)num}. See also the discussion at the end of Sec.\
\ref{bcs}.

\begin{figure}
\centering
\includegraphics[width=0.48\textwidth]{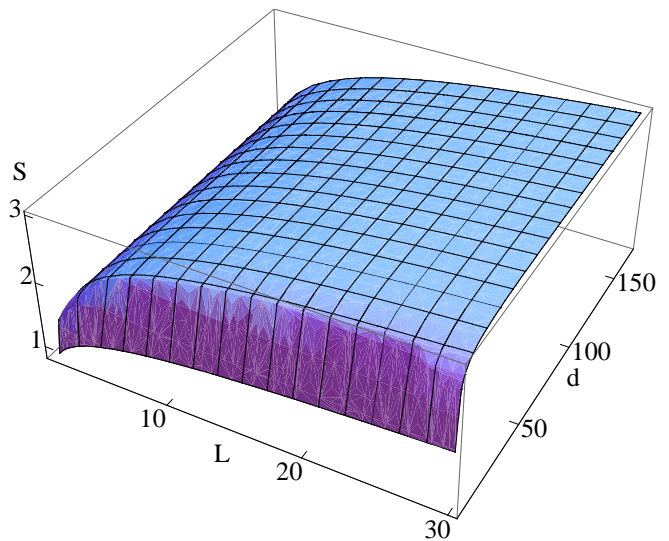}
\caption{(Color online)
Critical entropy ($\lambda=1$) between two blocks of $L$ spins at a
distance $d$ and the remaning part of the (infinite) chain. }
\label{fig:entropya}
\end{figure}

\begin{figure}
\centering
\includegraphics[width=0.48\textwidth]{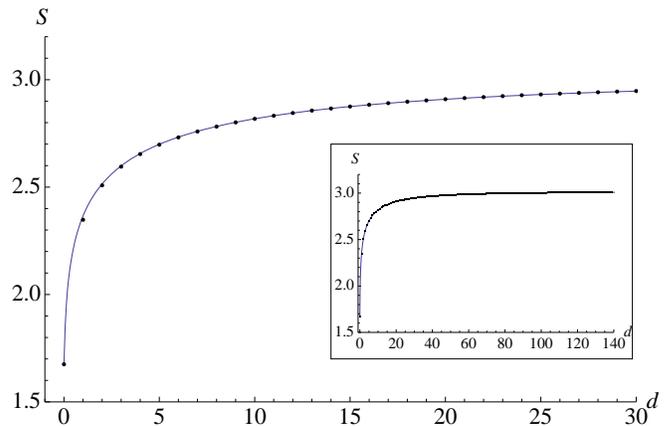}
\caption{(Color online)
Critical entropy ($\lambda=1$) between two blocks of $L=30$ spins and the
remaning part of the chain versus $d$. Inset: $L=30$ with $0<d<140$.
}
\label{fig:plotfit30inset}
\end{figure}

The global behavior of
\begin{eqnarray}
S(L,d)&=&\frac{1}{6}\Big(2\log(L-\alpha)-2\log(L+d) \nonumber\\
& & + \log(2L+d-\alpha) +\log(d+\alpha)\Big)
\nonumber\\
&=& \frac{1}{6}\Big(2\log L-2\log(L+d) \nonumber\\
& & + \log(2L+d) +\log d \Big) +\mathcal{K}(L),
\label{eq:entropy2}
\end{eqnarray}
with $\alpha$ and $\mathcal{K}(L)$ given by (\ref{eq:alpha}) and
(\ref{eq:KLexp}), respectively, is displayed in Fig.\
\ref{fig:entropya}.
The fit is accurate up to one part in $10^{3}$ for small $L$ ($<10$)
and one part in $10^{6}$ for $L>10$. Notice the logarithmic $L$
dependence for $d=0$ and the saturation effect for $L/d\ll1$. A
section of Fig.\ \ref{fig:entropya} is displayed in Fig.\
\ref{fig:plotfit30inset}. In particular, the inset shows
the asymptotic behaviour of the entropy and its saturation.

\subsubsection{Behavior of the critical entropy for small $d$}
\label{entsmalld}

Both in Figs.\ \ref{fig:entropya} and \ref{fig:plotfit30inset}
one notices for all values of $L$ a sharp entropy increase at small
values of $d$ from $d=0$ to $d=1$. This corresponds to the two
ordinates of the cusps in Fig.\ \ref{S:lambda}. Let us endeavor to
interpret this phenomenon on the basis of the formulas derived in
this section. Equation (\ref{eq:entropy1}) yields
\begin{eqnarray}
\Delta S &=& S(L,1) - S(L,0) \nonumber \\
&=& \frac{1}{6}\left(-2\log \frac{L+1}{L} + \log \frac{2L+1-\alpha}{2L-\alpha} + \log \frac{1+\alpha}{\alpha}
\right)  \nonumber \\
&\stackrel{L \gg 1 }{\sim}& \frac{1}{6}\log \frac{1+\alpha}{\alpha}
\nonumber \\
&\simeq& -\frac{1}{6}\log \alpha =\mathcal{K}
\label{eq:largederiv}
\end{eqnarray}
where we used Eq.\ (\ref{eq:alpha}) in the last equalities.
This agrees very well with Fig.\ \ref{fig:plotfit30inset} and explains why
$\Delta S$ is largely independent of $L$ in Fig.\ \ref{fig:entropya}.
More to this, the final result in Eq.\ (\ref{eq:largederiv}) yields a
suggestive interpretation of the fitting parameter
$\alpha$ in Eq.\ (\ref{eq:alpha}) and of the constant $\mathcal{K}$:
they turn out to be related to the entropy increase $\Delta S$ associated
with the opening of a $d=1$ gap (one qubit) in an interval of $2L$ contiguous spins.
It is therefore not surprising that
$\mathcal{K} = -(1/6) \log \alpha$, being $O(\Delta S)$, be also necessarily of order 1.

\section{conclusions}
\label{concl}

We provided an analytic and numerical treatment of the entanglement
entropy of two disjoint blocks of spins as a function of their
length and distance in the quantum Ising model with a transverse
magnetic field. We gave an analytic expression of the entropy at the
critical point, for two blocks of 1 and 2 spins at a generic
distance. We showed that the presence of a gap always
yields an entropy increase. At criticality, due to a logarithmic correction to the area law, this increase is less than a factor 2 for all values
of $d$, and becomes 2 for large $d$, when asymptotic additivity takes place.
We also showed that, interestingly, the entropy of the two blocks can be
written in terms of the entropy of a single block of spins, that, at
the quantum phase transition, grows logarithmically with the size of
the block. We have also given an accurate idea of the general
features of the entropy of two blocks as a function of their size
and distance.

The behavior of the entanglement in a critical spin chain agrees
with well known results in conformal field theory, where the
geometric entropy (analogous to the spin block entropy, but defined
in the continuum) can be computed for $1+1$ dimensional theories
\cite{Callan, FWilczek}.
The translation of field theoretical methods and ideas in the
language of quantum information will hopefully enable us to make use
of additional results for an arbitrary number of disjoint intervals
\cite{Cardy&Calabrese,Cardy&Calabrese2}.
This would of great interest from the point of view of multipartite
entanglement. The study of the statistical distribution of bipartite
entanglement for different bipartitions \cite{FFP} is a useful tool
for the analysis of multipartite entanglement. A deeper
comprehension of the dependence of entropy on distances and sizes of
blocks of spins could yield information about the role of quantum
phase transitions in the generation of multipartite entangled states
\cite{costantinietal,MMES}.


\acknowledgments We thank Matteo Paris for
interesting discussions. This work is partly supported by the
European Community through the Integrated Project EuroSQIP.

\appendix

\section{Computation of $S_L$}\label{computationS}

We review here the computation of the entropy of the reduced density
matrix $\varrho_L\equiv\Tr_{\neg L}\ketbra{\psi_0}{\psi_0}$ for $L$
adjacent spins  \cite{LRV,VLRK}. In the limit of infinite chain $(N
\to \infty)$, a given finite section of the chain is fully translational
invariant and $\varrho_L$ describes the state of any block of $L$
contiguous spins. The density matrix $\varrho_L$ can be
reconstructed from the restricted $2L\times 2L$ correlation matrix
$\Gamma_L ^A$ of Eq.\ (\ref{Gamma_A}).
In particular, a direct way to compute the spectrum of $\varrho_L$ and its entropy
$S_L$ from $\Gamma_L ^A$ is the following.

The matrix $\Gamma_L ^A$ can be put into a block-diagonal form by an
orthogonal transformation and its eigenvalues are purely imaginary
and come in pairs, $ \pm i \nu_l$,  and $|\nu_l|\leq 1$, with $1\leq
l
\leq L$. Let $V \in SO(2L)$ be the special orthogonal matrix such
that $\Gamma_L ^C = V\Gamma_L ^A V^T$ is block-diagonal
\begin{eqnarray}
\Gamma_L ^C =\bigoplus _{l=1} ^L \nu_l \left [ \begin{array}{cc} 0 & 1 \\
-1 & 0 \end{array} \right ].
\end{eqnarray}
Then, $V$ defines a new set of Majorana operators,
$\check{c}_m ^\dagger =\check{c}_m$,
\begin{equation}
\check{c}_m \equiv \sum_{n=1} ^{2L} V_{m,n} \check{a}_n,
\end{equation}
that satisfy the  same  anticommutation relations as the $\check{a}_n$'s,
and have correlation matrix  $\Gamma_L ^C$.
The structure of $\Gamma_L ^C$ implies that mode $\check{c}_{2l-1}$
is only correlated to mode $\check{c}_{2l}$. In the language of
fermionic operators, one gets $L$ spinless fermionic modes
\begin{eqnarray}
& & \hat{c}_l \equiv \frac{\check{c}_{2l-1} + i\check{c}_{2l}}{2},
\nonumber \\
& & \{\hat{c}_l,\hat{c}_m\} = 0, \quad \{\hat{c}_l ^\dagger ,
\hat{c}_m \} =\delta_{m n},
\end{eqnarray}
that, by construction, fulfill
\begin{equation}
\media{\hat{c}_l\hat{c}_m}=0, \qquad \media{\hat{c}_l ^\dagger \hat{c}_m}=\delta_{l m}\frac{1+\nu_l}{2}.
\end{equation}
Thus, the $L$ (nonlocal) fermionic modes are \emph{uncorrelated}, so that the reduced density matrix can be written as a product
\begin{equation}
\varrho_L = \rho_1 \otimes \ldots \otimes \rho_L.
\end{equation}
Now, the density matrix $\rho_l$, $1 \leq l \leq L$, has eigenvalues $(1\pm \nu_l)/2$
and entanglement entropy
\begin{equation}\label{entropy}
S(\rho_l)=H\left(\frac{1+\nu_l}{2}\right),
\end{equation}
where $H$ is the Shannon entropy of a bit (\ref{eq:Shannon}).
Therefore the spectrum of $\varrho_L$ results from the product of the spectra
of the density matrices $\rho_l$, and the entropy of $\varrho_L$ is
the sum of the entropies of the $L$ uncorrelated modes,
\begin{equation}\label{def:S_LA}
S_L =\sum_{l=1} ^L H\left(\frac{1+\nu_l}{2}\right) .
\end{equation}
This is Eq.\ (\ref{def:S_L}) of the text. Summarizing, for arbitrary
values of the magnetic field $\lambda$ and in the thermodynamic
limit, $N\rightarrow\infty$, the block entropy $S_L$ of the ground
state of the Ising model is given by the sum (\ref{def:S_LA}), where
$\pm i \nu_l$ are the pairs of imaginary eigenvalues of the block
correlation matrix $\Gamma_L ^A$  of Eq.\ (\ref{Gamma_A}).


\section{eigenvalues of $\Gamma_{2,d}^A$}\label{eigenvalues}

The coefficients in Eq.\ (\ref{quartic}) are functions of the
distance $d$ between the blocks and read
\begin{widetext}
\begin{eqnarray}
p(d) &=& \frac{2^6 (3+2d)^4 (5+2d)^4}{\pi^2 A} \left[5303
+\frac{24314 B}{3}+
\frac{41528 B^2}{9} + \frac{10144 B^3}{9} + \frac{896 B^4}{9} \right], \nonumber \\
 q(d) &=& \frac{2^{12} (3+2d)^2 (5 +2d)^2}{ \pi^4 A}  \left [203297+ 391466 B
 + \frac{2841841 B^2}{9} \right.
 \nonumber\\
 & & \phantom{\frac{2^6 (3+2d)^4 (5+2d)^4}{\pi^2 A} [} \left.  \; + \frac{3652160 B^3}{27} + \frac{2617216  B^4}{81} + \frac{329984 B^5}{81} + \frac{17152 B^6}{81}   \right ],  \nonumber \\
 r(d) &=& \frac{2^{22}
 (2+d)^4}{\pi^6 A} \left [12271
 +\frac{ 68116 B}{3} +\frac{158795 B^2}{9}  +  \frac{198074 B^3}{27} + \frac{139000 B^4}{81} + \frac{17312 B^5}{81} + \frac{896 B^6}{81} \right ],
 \nonumber \\
s(d) &=& \frac{2^{32}}{3^4 \pi^8 A}
(1 + d)^4 (2 + d)^8 (3 + d)^4,
\end{eqnarray}
\end{widetext}
where $A = (1 +2d)^2 (3 + 2d)^6 (5 +2 d)^6 (7 +2d )^2$ and
$B=d(4+d)$. We obtain $\mu^2$ by solving (\ref{quartic}). The eight
eigenvalues of $\Gamma_{2,d}^A$ will be
$\pm \mu_k = \pm i\nu_k$, with $0\leq \nu_k \leq 1$, for
$k=1,2,3,4$.


\begin{thebibliography}{99}

\bibitem{nielsen}
 M.A. Nielsen and I.L. Chuang,
    {\it Quantum Computation and Quantum Information}
    (Cambridge University Press, Cambridge, 2000).

\bibitem{Osterloh} A. Osterloh, L. Amico, G. Falci and R. Fazio, Nature \textbf{416}, 608 (2002).

\bibitem{Osborne} T. J. Osborne and M. A. Nielsen, Phys. Rev. A \textbf{66}, 032110 (2002).

\bibitem{LRV} J. I. Latorre, E. Rico, and G. Vidal, Quant. Inf. and Comp. \textbf{4} 048 (2004).

\bibitem{sarorev} L. Amico, R. Fazio, A. Osterloh, and V. Vedral, Rev. Mod. Phys. \textbf{80}, 517 (2008).

\bibitem{prosen}
T.~Prosen, I.~Pizorn, Phys. Rev. Lett.
\textbf{101}, 105701 (2008); T.~Prosen,  New J. Phys.
\textbf{10}, 043026 (2008).

\bibitem{VLRK} G. Vidal, J. I. Latorre, E. Rico, and A. Kitaev, Phys. Rev. Lett. \textbf{90}, 227902 (2003).

\bibitem{Korepin} B.-Q. Jin and V.E. Korepin, J. Stat. Phys. \textbf{116}, 79 (2004).

\bibitem{Cardy&Calabrese} P. Calabrese and J. Cardy, J. Stat. Mech. P06002 (2004).

\bibitem{eisert}
J. E. Eisert and M. Cramer, Phys. Rev. A \textbf{72}, 042112
(2005).

\bibitem{Its} A. R. Its, B.-Q. Jin and V. E. Korepin, J. Phys. A, \textbf{38}, 2975 (2005).

\bibitem{cft}
P. Di Francesco, P. Mathieu, and D. Senechal, {\it Conformal Field Theory} (Springer,
Heidelberg, 1999).

\bibitem{sachdev} S. Sachdev, {\it Quantum Phase Transitions} (Cambridge Univ. Press, 1999).

\bibitem{Callan} C. G. Callan and F. Wilczek, Phys. Lett. B, \textbf{333} (1994).

\bibitem{FWilczek} C. Holzhey, F. Larsen, and F. Wilczek, Nucl. Phys. B \textbf{424} 44 (1994).

\bibitem{Casini} H. Casini and M. Huerta, Phys. Lett. B \textbf{600}, 142 (2004).

\bibitem{Lieb} E. Lieb, T. Schultz and D. Mattis, Annals of Phys. \textbf{16},
407 (1961).

\bibitem{Katsura} S. Katsura, Phys. Rev. \textbf{127} 1508 (1962).

\bibitem{Barouch1} E. Barouch, B. McCoy, and M. Dresden, Phys. Rew. A \textbf{2} 1075 (1970).

\bibitem{Barouch} E. Barouch and B. McCoy, Phys. Rev. A \textbf{3} 786 (1971).

\bibitem{Cardy&Calabrese2} P. Calabrese and J. Cardy, J. Stat. Mech. P04010 (2005).

\bibitem{CamposVenuti} L. Campos Venuti, C. degli Esposti Boschi, M. Roncaglia, and A. Scaramucci, Phys. Rev. A, \textbf{73}, 010303(R) (2006).

\bibitem{RigolinPRA} G. Rigolin, T. R. de Oliveira, and M. C. de Oliveira, Phys. Rev. A \textbf{74}, 022314 (2006).

\bibitem{RigolinPRArapid} T. R. de Oliveira, G. Rigolin, and M. C. de Oliveira, Phys. Rev. A
\textbf{73}, 010305(R) (2006).

\bibitem{RigolinPRL} T. R. de Oliveira, G. Rigolin, M. C. de Oliveira, and E. Miranda, Phys. Rev. Lett.
\textbf{97}, 170401 (2006).

\bibitem{Keating} J.P. Keating, F. Mezzadri, and M. Novaes, Phys. Rev. A \textbf{74}, 012311 (2006).

\bibitem{arealawrev} J. Eisert, M. Cramer, and M.B. Plenio, {\it Area laws for the entanglement entropy - a review}, Preprint arXiv:0808.3773 (2008).

\bibitem{FFP}
P. Facchi, G. Florio and S. Pascazio, Phys. Rev. A \textbf{74},
042331 (2006).

\bibitem{costantinietal}
G. Costantini, P. Facchi, G. Florio, and S. Pascazio, J. Phys. A:
Math. Theor. \textbf{40}, 8009 (2007).

\bibitem{MMES}
P.~Facchi, G.~Florio, G.~Parisi, S.~Pascazio,  Phys. Rev. A
\textbf{77}, 060304(R) (2008).



\end{thebibliography}
\end{document}